**Title**

# The role of wild birds in the global highly pathogenic avian influenza H5 panzootic

**Authors**: Manon COUTY[1], Claire GUINAT[1], Diletta FORNASIERO[2], François-Xavier BRIAND[3], Pierre-Yves HENRY[4,5], Béatrice GRASLAND[3], Loïc PALUMBO[6], Guillaume LE LOC'H[1]

**Affiliations**:

[1] IHAP, Université de Toulouse, INRAE, ENVT, Toulouse, France

[2] Istituto Zooprofilattico Sperimentale delle Venezie, Legnaro, Italy

[3] National Agency for Food Environmental and Occupational Health and Safety, Ploufragan, France

[4] Mécanismes adaptatifs et évolution (MECADEV UMR 7179), Muséum National d'Histoire Naturelle, Centre National de la Recherche Scientifique, Brunoy, France

[5] Centre de Recherches sur la Biologie des Populations d'Oiseaux (CRBPO), Centre d'Ecologie et des Sciences de la Conservation (CESCO UMR 7204), Muséum National d'Histoire Naturelle, Centre National de la Recherche Scientifique, Sorbonne Université, Paris, France

[6] French Biodiversity Agency, Orléans, France

**Abstract**

The highly pathogenic avian influenza (HPAI) H5 clade 2.3.4.4b has triggered an unprecedented global panzootic. As the frequency and scale of HPAI H5 outbreaks continue to rise, understanding how wild birds contribute to shape the global virus spread across regions—affecting poultry, domestic and wild mammals—is increasingly critical. In this review, we examine ecological and evolutionary studies to map the global transmission routes of HPAI H5 viruses, identify key wild bird species involved in viral dissemination, and explore infection patterns, including mortality and survival. We also highlight major remaining knowledge gaps that hinder a full understanding of wild birds' role in viral dynamics, which must be addressed to enhance surveillance strategies and refine risk assessment models aimed at preventing future outbreaks in wildlife, domestic animals and safeguard public health.

## Introduction

The highly pathogenic avian influenza (HPAI) panzootic caused by H5 viruses of clade 2.3.4.4b is unprecedented in scale and impact. Since 2020, HPAI H5 viruses have spread to numerous countries worldwide, reaching over 76 countries across Europe, Asia, Africa, the Americas[1], and even Antarctica[2], causing massive die-offs in wild birds, severe outbreaks in poultry, and increasing spillovers to mammals[3–6]. While this shift in viral dynamics is recent, the history of avian influenza viruses extends far into the past.

Historically, avian influenza viruses have been naturally maintained in wild waterbirds, particularly those in the orders Anseriformes and Charadriiformes[7], as low pathogenic forms, with HPAI cases reported only sporadically in wild birds[8]. HPAI outbreaks were mostly confined to poultry, typically arising when low pathogenic forms introduced from wild waterbirds evolved into highly pathogenic forms[9]. However, the emergence of the A/Goose/Guangdong/1/1996 (Gs/Gd) HPAI H5 lineage in 1996 in China marked a critical turning point. This lineage became established in poultry, and spillovers to wild bird populations were first reported in 2002[10]. From 2005, wild birds cases became more frequent, spreading across Asia, Europe and Africa[11,12]. Over time, the Gs/Gd HPAI H5 lineage diversified into multiple clades, with clade 2 increasingly detected in wild birds[13–15]. In 2014, the clade 2.3.4.4 spread beyond Asia[16], with migratory wild birds playing a key role in spatial dissemination[16–18].

Since 2020, the spread of HPAI H5 clade 2.3.4.4b has escalated into a global panzootic[19–21]. While circulating in Eurasia and Africa in 2020, HPAI H5 viruses surged in North America in 2021 and reached South America in 2022[22,23]. The 2020-2021 wave was primarily driven by the H5N8 subtype[24,25] with a peak in wild bird detections in winter followed by a sharp decline in summer months (Fig. 2A, Extended Data Fig. 1)[24]. The H5N1 subtype completely dominated in subsequent waves[25,26] and persistent viral circulation in summers 2022 and 2023 was observed, differing from the usual HPAI seasonality and affecting a wide spectrum of hosts, notably colonial breeding birds. Globally, reported cases have encompassed over 500 wild bird species spanning 25 orders, with more than half of the species never reported as infected before 2021[27]. The death toll on wild birds has been particularly high, though difficult to quantify. Some species, previously spared, have experienced mass mortality events, with potential long-term consequences. The ongoing viral circulation in wild birds has also fueled large-scale outbreaks in poultry[1,28] and contributed to repeated spillovers in wild and domestic mammals[3–6], highlighting the role of wild birds in cross-species transmission. The rising incidence of HPAI H5 viruses in mammals is particularly alarming[29], as it raises concerns about viral adaptation to these hosts, posing risks to public health.

Wild birds have emerged as both victims and vectors of HPAI H5 viruses, experiencing unprecedented mortality while driving the virus's spatial dispersion and seeding outbreaks in poultry and mammals. A better understanding of their role in this global panzootic is therefore essential to study these new viral dynamics and gain broader insights into HPAI H5 epidemiology. In this review, we systematically examine the ecological and evolutionary studies of HPAI H5 viruses in wild birds since 2020 to trace the global routes of viral dissemination. We identify key wild bird species involved, examining infection patterns, mortality and asymptomatic cases, as well as transmission pathways linked with their ecological behaviors. Finally, we identify major research gaps that hinder a full understanding of wild birds' role in HPAI dynamics - gaps that must be addressed to improve effective surveillance and prevention strategies.

## Global dissemination routes of HPAI H5 viruses via wild birds

### *Spatial spread in Eurasia in 2020-2023*

In summer 2020, HPAI H5 viruses were detected in North Asia (Russia, Kazakhstan)[30,31], with viruses likely originating from North Africa or the Middle East[25,32,33] (Fig. 1). North Asia has likely been acting as a key viral hub, due to its role as a major breeding ground for many migratory bird populations and its location at the intersection of multiple flyways[33,34]. Phylogeographic analyses indicated that by autumn 2020, viruses had disseminated widely from North Asia to Central Asia (China)[30,31,33,35–38], East Asia (Japan and South Korea)[33,39], South Asia[38] and Europe[32,40–42,30,43,44,39,38,31,33,34,25,44] (Fig. 1). Similar dissemination patterns from North Asia to East Asia[45–48] and Europe[25,49,50] were reported in autumn 2021 and 2022. Eastward viral movements from Europe to Asia were also documented, with viruses detected in North, Central and East Asia in autumn 2020 found closely related to

European ones[30,31,33,35–39,51–54], a pattern repeated in autumn 2021[36,54–57] and 2022[58] (Fig. 1). Frequent bidirectional viral circulation between Central and East Asia in 2020-2023 was also supported by phylogeographic analyses, potentially in both autumn and spring[30,31,34,37,38,59]. These extensive viral disseminations in Eurasia were closely and timely aligned with wild bird migrations[34,38,60]. In the Northern Hemisphere, migration is often categorized into autumn movements, when birds travel from breeding to more southerly wintering grounds, and spring movements when they return to breeding grounds. Birds engaging in these journeys while infectious have thus the potential to disseminate viruses along their routes.

Among wild birds, Anatidae, particularly dabbling ducks, have been considered as primary drivers of viral spread in Eurasia. Movement tracking of infected mallards (*Anas platyrhynchos*) in China showed it was capable of movements while infected[61] and other individuals were found alive while infected across Eurasia, with or without clinical signs[62–67]. Other species found alive while infected in Eurasia included Eurasian wigeons (*Mareca penelope*)[63–65,67–69], Eurasian teals (*Anas crecca*)[62,67,69,70], spot-billed ducks (*Anas poecilorhyncha*), northern pintails (*Anas acuta*), falcated teals (*Mareca falcata*)[62] and northern shovelers (*Spatula clypeata*)[69]. In East Asia, phylodynamic analyses showed whooper swans (*Cygnus cygnus*) could have played an important role in viral introductions[30,47], while phylogeny suggest hooded cranes (*Grus monacha*) could be involved in regional dissemination[71]. Geese also appeared involved in viral movements in Europe as barnacle *(Branta leucopsis)*, greylag (*Anser anser*), and pink-footed (*Anser brachyrhynchus*) geese were alive while infected, with or without clinical signs[63–65,67].

Following viral introductions from North Asia into Europe, cases in migratory Anatidae surged around October of each year[32,40,49,50]. Cases were particularly clustered along the coasts of the Wadden and Baltic seas in Germany, the Netherlands, the United Kingdom (UK) and Denmark[32,40,49,50], though dependent on surveillance efforts. This region seemed important for further dissemination across Europe in 2020-2022[25,32,34,39,44,72,73], while extensive viral diffusion was observed with cases reported in Iceland, Svalbard, and Jan Mayen islands for the first time in 2022[50,73].

### ***Intercontinental spread to North America in 2021-2022***

In December 2021, HPAI H5 viruses were detected in Canada[22]. Phylogeographic analyses revealed multiple introductions from Europe across the Atlantic Ocean in 2021[25,34,54,59,72,74,75] and 2022[75–77] (Fig. 1), probably occurring in two stages. First, birds migrating from European wintering grounds to northern breeding grounds (Iceland, Greenland) may have carried the virus[22,73,76]. Transmission could then have occurred to birds breeding in these areas but wintering in North America. Several species could have played a role in this intercontinental spread, at least partially, including Eurasian wigeons, barnacle geese, great skuas (*Stercorarius skua*) or black-headed gulls (*Chroicocephalus ridibundus*), that were notably found as vagrants from Europe in Canada[22]. In parallel, multiple viral introductions occurred on the Pacific coast of North America in 2022, likely originating from East Asia, supported by phylogeographic analyses[25,75,78–80]. Bidirectional viral flow through the Pacific seemed to occur as viruses detected in East Asia in autumn 2022 were closely related to those in North America[46]. Various species moving between both continents were suggested as possible vectors, although evidence remains limited[79].

Following introduction, HPAI H5 viruses rapidly spread across North America (Fig. 1)[19,72,74,75,79,80], mostly those introduced on the Atlantic coast, with Anatidae considered primary drivers. Movement tracking of infected mallards and lesser scaups (*Aythya affinis*) showed their ability to move while infected[81,82]. Other species found alive while infected include mallards, American green-winged teals (*Anas crecca carolinensis*), American wigeons (*Mareca americana*), northern pintails, American black ducks (*Anas rubripes*), cackling geese (*Branta hutchinsii*) and snow geese (*Anser caerulescens*)[74,80,83,84]. A spatial transmission modelling study also suggested that other species, like pelicans, may have contributed to viral spread[85].

### *Dissemination to Central and South America in 2022-2023*

From late 2022, HPAI H5 viruses reached Central and South America via multiple independent viral introductions from North America to Colombia, Peru, Ecuador and Venezuela, as shown by phylogeographic analyses[23,86–89] (Fig. 1). These introductions coincided with autumn migrations to South American wintering grounds[90], although the specific species involved remain unclear. The introduction to Peru was particularly impactful, leading to regional spread within Peru and Chile and mass mortality in wild birds and marine mammals[23,91–93]. Subsequent viral dissemination occurred in 2023 to Uruguay[88,94,95], Argentina[6,96,97] and Brazil[91,98,99], as shown by phylogeographic analyses or phylogeny. By late 2023, HPAI H5 viruses have spread from South America to the subantarctic region[2,100,101]. Brown skuas (*Stercorarius antarcticus*) and giant petrels (*Macronectes* spp.) were among the first affected species and could have contributed to the introduction in this region[100,101]. Other seabird species with migratory connectivity to the South American coast may also have played a role[102].

### *Circulation in Africa in 2020-2023*

HPAI H5 viruses were also reported in Africa in 2020-2023, although available data remain limited. Initial viral introductions were detected in West Africa, notably in Senegal and Nigeria in late 2020, likely originating from Europe via autumn migration, as shown by phylogeographic analyses (Fig. 1)[25,44,54,103,104]. In Nigeria, white storks (*Ciconia ciconia*) and ruffs (*Calidris pugnax*), both migrants from Europe to Africa, were found healthy while infected, suggesting their possible role in viral introduction[104]. Viruses detected in wild birds in Egypt in 2021 were also closely related with those circulating in Europe[105].

Evidence of viral dissemination within Africa has been observed, as viruses detected in southern Africa and Nigeria in 2021-2022 were genetically linked to those from Senegal in late 2020 and early 2021[103,104]. In Nigeria, intra-African migratory species such as African jacanas (*Actophilornis africanus*), and resident birds like white-faced whistling ducks (*Dendrocygna viduata),* spur-winged geese (*Plectropterus gambensis*) and square-tailed nightjars (*Caprimulgus fossii*) were found healthy while infected, indicating potential for local viral dissemination[104]. Backflow of viruses from Africa to Europe also likely occurred via spring migration in 2020-2023, as shown by phylogeographic analyses, although less documented (Fig. 1)[54,59,106].

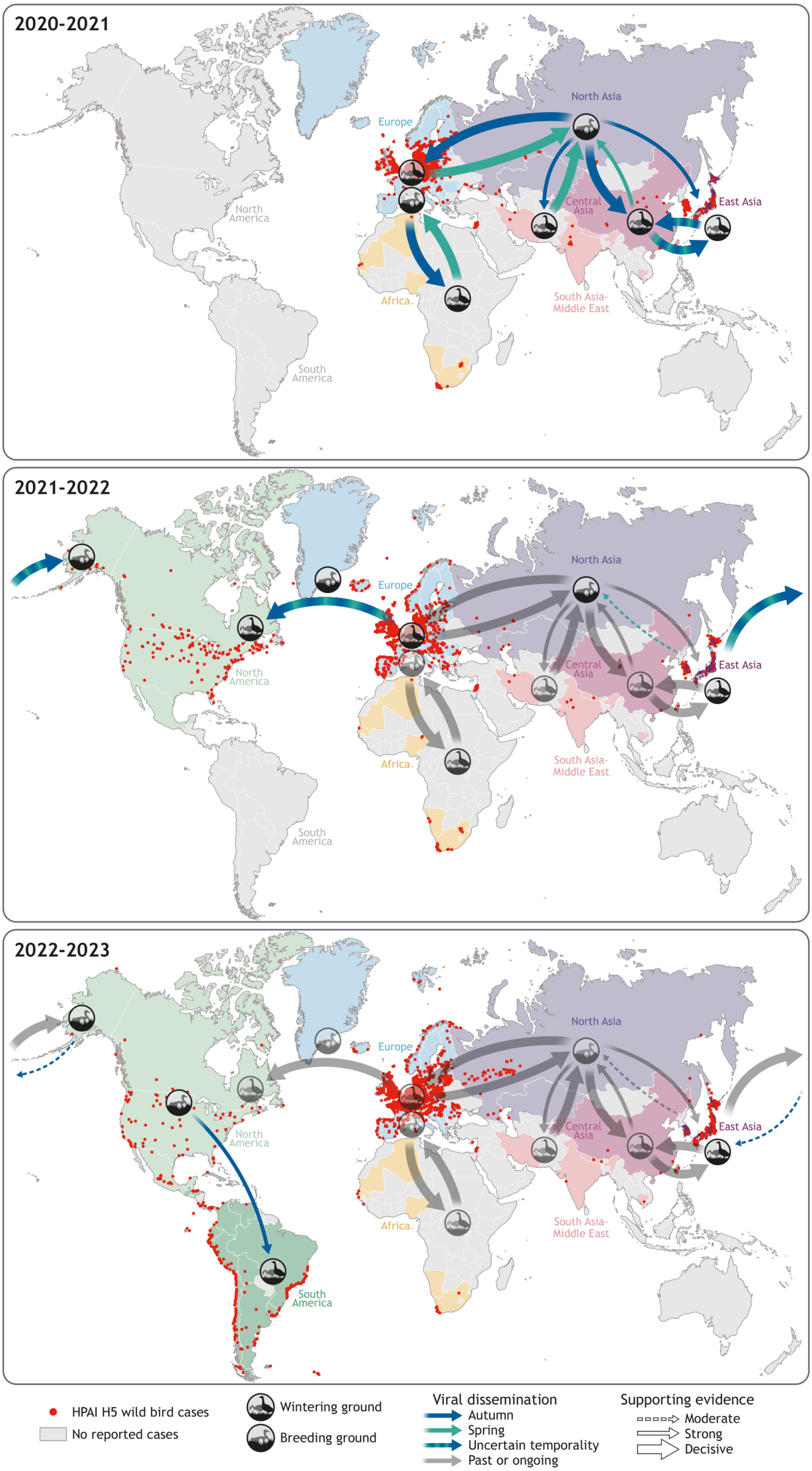

**Figure 1. Global dissemination routes of HPAI H5 viruses via wild birds from 2020 to 2023.** Red dots represent HPAI H5 cases (H5Nx, H5N8, H5N1) in wild birds, as reported to World Animal Health Information System (WAHIS) of the World Organization for Animal Health. A case has been defined as a reported case in a wild bird species, per location and per month, as reported to WAHIS. Each map shows an epidemiological season, defined from October 1st to September 30$^{th}$ (extended to the end of 2023 for the 2022-2023 wave). Regions were defined on epidemiological, geographical and ecological criteria. Arrows represent viral dissemination linked to wild bird movements based on the systematic literature review, with shape showing different levels of evidence based on phylogenetic (moderate) or phylogeographic analysis (strong support: Bayes Factor (BF) <100; decisive support: BF>100). Links to bird migratory networks were made based on temporality as well as ecological criteria. Black icons indicate major breeding and wintering grounds shared by various species. Grey arrows represent previously observed viral movements that could still be ongoing in the current season.

## Impacts of HPAI H5 viruses on wild bird populations

### *Unprecedented mortalities in colonial breeding birds*

HPAI H5 viruses have caused unprecedented mortality in colonial breeding birds, particularly from summer 2022 with many species reported infected for the first time[27]. Quantifying the full impact is challenging, as a single reported case could represent thousands of infected birds (Fig. 2A, Table 1), and many events likely went unreported. Among seabirds, northern gannets (*Morus bassanus*) were among the most severely affected species[65], with approximately 75% of North Atlantic colonies impacted in summer 2022[107] (Fig. 2F, 2G, Table 1). In Europe, several colonies were affected including Scotland, with a 70% reduction in colony size and a 75% decline in breeding success[107], and France, with adult mortality exceeding 58% and 67% for chicks[108]. In North America, 11.5% of the population was lost with reproductive success dropping to 17% in one colony[109–111]. Russian colonies also underwent a 40% decline in inhabited nests[112]. Despite these severe losses, recovery signs have emerged, with birds developing antibodies and iris color changes potentially indicating prior infection[107,108,112,113]. Other seabird species experienced significant mortality (Table 1), like great skuas, in Great Britain in summer 2021 and 2022, with up to 11% breeding population decline[114]. Common murres (*Uria aalge*), Atlantic puffins (*Fratercula arctica*) and razorbills (*Alca torda*)[109,110,115] were affected in North America, while Peruvian boobies (*Sula variegata*)[86,89,92,116–118], Humbolt penguins (*Spheniscus humboldti*)[119], and magnificent (*Fregata magnificens*) and great frigatebirds (*Fregata minor*)[120] suffered losses in South America. Snowy albatrosses (*Diomedea exulans*), gentoo penguins (*Pygoscelis papua*) and brown skuas were affected in the Antarctic region[2,100,101], with recoveries after infection observed for this species. In South Africa, African penguins (*Spheniscus demersus*) were also notably impacted[103].

Laridae species, including terns and gulls, also experienced devastating outbreaks (Table 1). Sandwich terns (*Thalasseus sandvicensis*) were particularly affected in summer 2022, with outbreaks in 60% of European colonies, leading to an estimated 17% loss of the total breeding population[121] (Fig 2D, Table 1). Mortality rates exceeded 30% in some German colonies[122], while some Dutch colonies saw up to a loss of all breeding adults[123]. Common terns (*Sterna hirundo*) also had severe losses in summer 2022 in Germany[122], Great Britain[124], Canada[109] and Ireland in summer 2023[125] (Fig 2B, 2C, Table 1). Caspian terns (*Hydroprogne caspia*) lost over 60% of the Lake Michigan population in the United States (US) in summer 2022[19] and more than half a colony on the Pacific coast in summer 2023[126]. Other affected tern species included Arctic (*Sterna paradisaea*), roseate (*Sterna dougallii*)[125], swift (*Thalasseus bergii*)[127] and South American terns (*Sterna hirundinacea*)[6]. Among gulls, herring gulls (*Larus argentatus*) were affected in Europe in 2022[65] while black-headed gulls became particularly affected from early 2023[26,106,128–130]. Other gulls with significant mortality included great black-headed (*Ichthyaetus ichthyaetus*), Caspian (*Larus cachinnans*)[58], great black-backed (*Larus marinus*)[131] and kelp gulls (*Larus dominicanus*)[116,118] as well as black-legged kittiwakes (*Rissa tridactyla*)[109,110] (Fig. 2H, Table 1). The emergence of a new HPAI H5N1 genotype, which likely emerged from a reassortment with a gull-adapted low pathogenic virus, may have contributed to the high severity of populational impacts[25,129]. Differential survival between adults and chicks in gulls[124,126] and terns[121,127], suggests possible acquired immunity, supported by H5 antibodies detection in herring gulls[132]. In summer 2023, Sandwich terns in the UK showed low mortality, also pointing out possible immune protection[125].

Several waterbird species experienced significant population losses in 2020-2023 (Table 1). Dalmatian pelicans (*Pelecanus crispus*) lost around 60% of a colony in Greece in early 2022[133,134] with additional die-offs in Russia[70]. Peruvian pelicans (*Pelecanus thagus*) experienced significant mortality along the coasts of Peru and Chile from late 2022[23,86,87,89,92,116–118], while great white pelicans (*Pelecanus onocrotalus*)[44,135,136] and American white pelicans (*Pelecanus erythrorhynchos*)[19,83] were also impacted. Cormorants were also heavily affected, like Cape cormorants (*Phalacrocorax capensis*) that lost an estimated 30% of their population in Southern Africa[103,137]. Guanay cormorants (*Leucocarbo bougainvilliorum*) suffered heavy losses in Peru and Chile[92,116,118], double-crested cormorants (*Nannopterum auritum*) in North America[19,83,109], and great cormorants (*Phalacrocorax carbo*) in the Baltic region, though with probable limited population-level impacts[132]. In cranes, hooded cranes (*Grus monacha*) lost about 10% of their population in Japan and Korea[71] while common cranes (*Grus grus*) suffered high losses in Israel[135]. Black necked grebes (*Podiceps nigricollis*) also experienced significant mortality in China[138,139] and Canada[83].

***Mortalities in other wild bird groups***

Other wild bird groups, including Anatidae, raptors, and land birds, were also affected with many species reported infected for the first time. Mortality per outbreak was generally lower than in colonial breeding birds due to less gregarious behaviors, but impacts could still have been substantial for long-lived species like raptors.

Anatidae showed the most reported cases (Fig. 2A), with mute swans (*Cygnus olor*) and geese, especially barnacle and greylag geese among the most reported species in Europe in 2020-2023 with more than 850, 950 and 650 reported cases respectively[63–65,131,140–144]. In the UK, a massive die-off of barnacle geese caused a 32% population loss, while a high percentage of survivors developed antibodies[145]. In North America, snow geese were particularly affected with more than 18,000 cases reported in 2021-2023, and so were Ross's (*Anser rossii*) and Canada *(Branta canadensis*) geese with around 950 and hundreds of reported cases, respectively[19,75,83,110,146]. Among dabbling ducks, Eurasian wigeons and mallards were frequently reported in Europe[32,40,49,143], while lesser scaups were notably affected in the US with around 1,500 reported cases[19]. In North America, common eiders (*Somateria mollissima)*, a colonial sea duck considered near threatened, experienced severe impacts on the Atlantic coast in summer 2022, with mortality rates reaching up to 18% in a single colony[83,109,147].

Raptors were also significantly affected, with common buzzards (*Buteo buteo*) and peregrine falcons (*Falco peregrinus*) among the most reported raptor species in Europe, with more than 300 and 70 reported cases, respectively[63–65,131]. White-tailed sea eagles (*Haliaeetus albicilla*) also experienced notable losses in Norway and Germany[148,149], while bald eagles (*Haliaeetus leucocephalus*) were affected in North America with more than 400 cases[19,75,150]. Colonial raptors were also affected, including griffon vultures (*Gyps fulvus*) in France and Spain[151] (Fig. 2E), turkey vultures (*Cathartes aura*) in Chile[86,118], and black vultures (*Coragyps atratus*) in North America with more than 1,900 cases[19,75]. In griffon vultures, immobility was observed, with high juvenile mortality, while many adults survived and developed antibodies[151]. Great horned owls (*Bubo virginianus*) were also affected and some showed signs of recovery[75,152].

Among land birds, corvids were particularly impacted. Hundreds of large-billed crows (*Corvus macrorhynchos*) died in Japan[153]. Similarly, American crows (*Corvus brachyrhynchos*) were highly affected[75], with over 500 cases reported in Canada[110]. Mortality was also observed in game birds, including common pheasants (*Phasianus colchicus)* in Finland[154] and wild turkeys (*Meleagris gallopavo*) in the US[155].

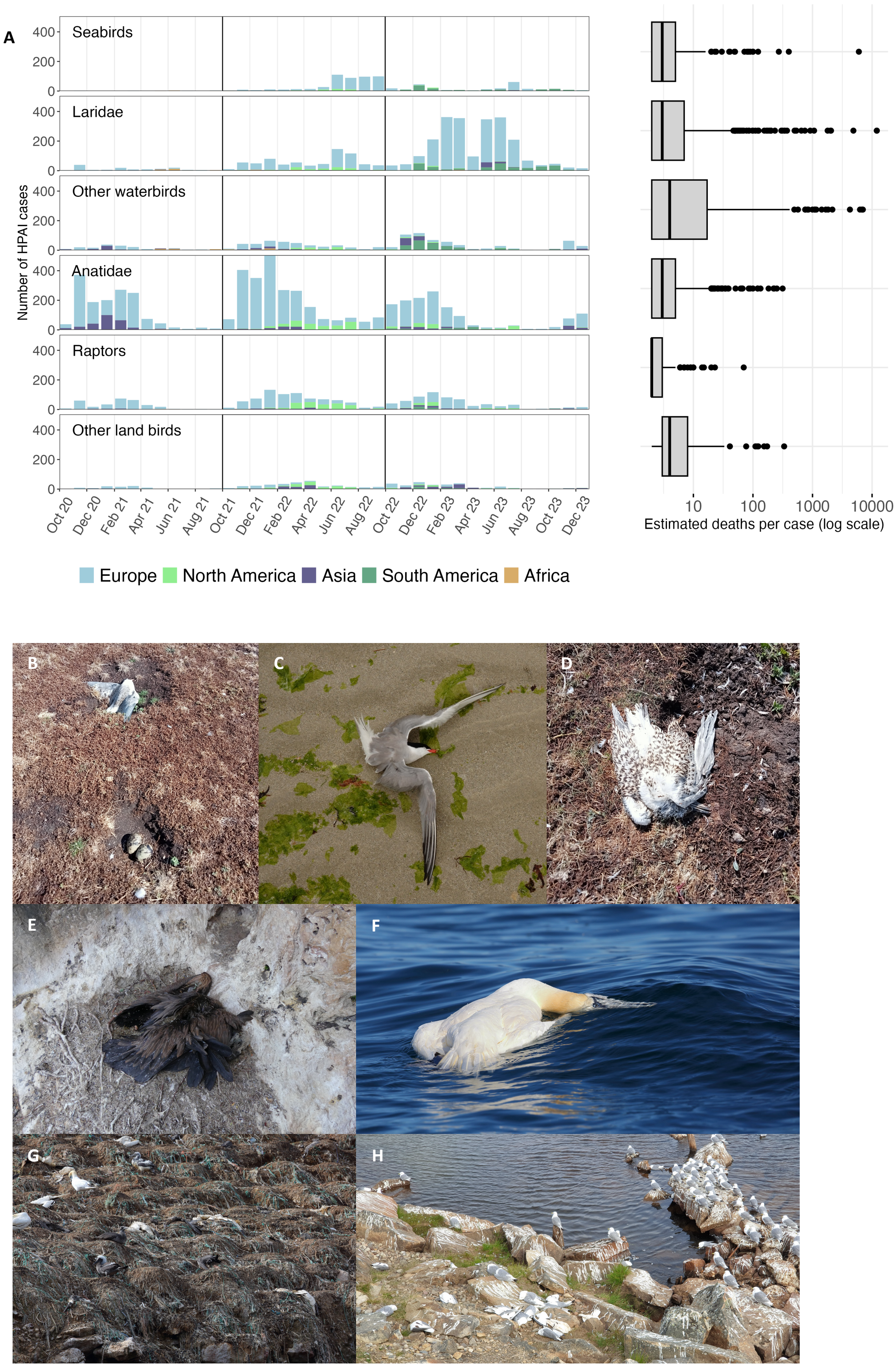

**Figure 2. Impact of HPAI H5 viruses in wild bird populations from 2020 to 2023.** Top panel: A. Bar plots represent the temporal distribution of HPAI H5 cases (H5Nx, H5N8, H5N1) in wild birds, as reported to World Animal Health Information System (WAHIS) of the World Organization for Animal Health. A case has been defined as a reported case in a wild bird species, per location and per month, as reported to WAHIS. Seasons were defined from October 1st to September 30th (extended to the end of 2023 for the 2022-2023 wave). Species were categorized into six groups based on taxonomic, epidemiological, and ecological criteria: Seabirds (excluding Laridae), Laridae, Other waterbirds, Anatidae, Raptors, and Other land birds (Extended Data Table 1). The boxplot represents the estimated number of dead birds per case reported in WAHIS for each wild bird group. It shows that quantitative aspects of mortality are not well represented like in mass mortality events, where one case could truly represent thousands. Bottom panel: Pictures of wild birds affected by HPAI H5 viruses: B. Juvenile common tern with abandoned eggs in France in summer 2023 (Source: Bernard Cadiou), C. Common tern in France in summer 2023 (Source: Yann Jacob), D. Juvenile Sandwich tern in France in summer 2023 (Source: Bernard Cadiou), E. Griffon vulture in France in summer 2022 (Source: Robert Straughan), F,G. Northern gannets in France in summer 2022 (Source: Pascal Provost), H. Black-legged kittiwakes in Norway in summer 2023 (Source: Pierre-Yves Henry).

## Ecological drivers of HPAI H5 infections in wild birds

Identifying the mechanisms by which wild birds become infected with HPAI H5 viruses is often challenging. However, certain ecological behaviors, such as migrations, breeding and feeding, can help explain patterns of exposure and transmission. Migratory birds, especially Anatidae, have played a key role in the long-distance spread of HPAI H5 viruses[34,60,156]. Critical nodes along their migratory routes—breeding, wintering, and stopover grounds—likely facilitated virus transmission through increased interspecies contact, environmental contamination, viral circulation, and reassortments[30,53,61,62,69,71,81,133,138,157]. However, the significant impacts during summer months, when birds are largely sedentary, and occurrences in non-migratory species suggest that additional regional transmission mechanisms were also at play[158].

Certain species, such as gulls and skuas, may have acted as bridge hosts. Their opportunistic feeding or kleptoparasitic behaviors likely increased both their risk of exposure and their potential to disseminate the virus between colonies and into previously unaffected areas[100,101,107,108,121,124,159,160]. Connectivity between colonies likely further promoted viral dissemination, facilitating inter-colonial transmission in species such as Sandwich terns[121,123], great cormorants[132], great skuas[114], and gannets, with both breeding and non-breeding individuals observed visiting other colonies during outbreaks for this species[107,108,111–113,159].

Once introduced into dense breeding colonies, HPAI H5 viruses may have spread rapidly due to close contacts between individuals[122,124,126,131,133,145]. For example, ground-nesting cormorant colonies were significantly more affected than tree-nesting ones, likely due to their higher colony densities[132]. Decomposing carcasses further contributed to environmental contamination and may have acted as secondary sources of infection[122,145,161], as observed in terns and sanderlings[6,95].

Predation or scavenging of infected preys appeared to be a major infection route for raptors, given their dietary habits, and is supported by field observations, spatiotemporal clustering of cases and genetic analyses[47,71,84,92,132,145,146,148–150,152,153,162–165]. This route is also probable in predatory or scavenging seabirds, such as skuas, gulls, frigatebirds, albatrosses, and petrels[6,100,101,114,124,126,160], as well as for land birds, notably corvids[164–166].

## Discussion

### *Insights on the HPAI H5 panzootic*

We identified the major global transmission routes of HPAI H5 viruses via wild birds, which are critical for understanding viral ecology and anticipating where and when viruses might spread, thereby informing future risk assessments. These routes have shown strong alignment with migratory patterns[34,38,60], though predicting them remains challenging due to variability across species, seasons and regions. However, certain areas, including North Asia and Europe, have consistently emerged as key viral hubs for both maintenance and onward spread[17,18,72,167,168]. Proximity to major migratory flyways and critical habitats such as breeding, wintering and stopover grounds has been associated with increased risk of exposure[53,156,169]. These regions therefore represent valuable targets for enhanced surveillance and may serve as sentinel sites to monitor viral evolution, provided that species-specific migration routes and timings are taken into account[170].

Identifying the species responsible for both long-range and local spread remains complex. Dabbling ducks and other Anatidae species have frequently been identified as primary drivers of intercontinental spread[13–16,168]. Local maintenance and shorter-distance transmission may have involved other groups, such as gulls, whose roles remain less well defined. Continued targeted sampling of live birds is essential to detect asymptomatic carriers that may contribute to viral dissemination[170].

Importantly, this panzootic has affected an unprecedented diversity of wild bird species, including many not previously reported infected by HPAI H5 viruses[27], raising serious conservation concerns. Colonial breeding species, such as seabirds or terns, have suffered mass mortality events, resulting in sharp population declines and probable disruptions of colony social structures. Long-lived species with low reproduction rates, restricted geographic ranges or threatened status are particularly at risk, as outbreaks may have caused long-term effects on their populations. The breadth of affected species also raises broader ecological concerns, suggesting an expansion of the virus's host range and the emergence of novel wild bird reservoirs capable of sustaining viral circulation. This may help explain the unusual persistence of HPAI H5 viruses throughout the year in multiple regions—a major difference from patterns observed in previous epizootics—although the underlying mechanisms remain poorly understood[158]. Increased viral circulation has also led to repeated spillover events into poultry[1,28], and into both domestic and wild mammals[3–6]. Both terrestrial and marine mammals have been affected, including large-scale mortality events among marine mammals in South America[6,23,171]. Reports of probable mammal-to-mammal transmission raise additional concerns about the virus's adaptation to new hosts, including humans[3,4,6,23]. These developments underscore the urgent need for continued research and international collaboration to mitigate the impacts of HPAI H5 viruses on wildlife, poultry and public health, while enhancing preparedness for future epizootics.

***Challenges and future directions***

Despite the insights gained through this systematic literature review, many aspects of the epidemiology of HPAI H5 viruses and their relationship with wild bird ecology remain poorly understood. Key questions persist, such as the specific roles different species play in maintaining and disseminating the virus, why certain populations have experienced massive die-offs while others have been relatively spared or why some regions, like Oceania, have so far remained unaffected[172]. A major limitation lies in the nature of the available data. Current knowledge is largely derived from surveillance networks, which are highly variable in geographic coverage, temporal resolution, and methodological consistency. These networks are further shaped by factors including species detectability, habitat accessibility, conservation status, perceived epidemiological relevance, and levels of public awareness. As a result, substantial sampling biases exist both within and between countries, complicating efforts to interpret findings as representative of broader patterns.

Moreover, only a fraction of wild bird cases is reported in public databases, and reporting practices vary across regions due to inconsistent case definitions and diagnostic criteria. These discrepancies hinder cross-regional comparisons and contribute to an incomplete understanding of outbreak extent and severity. Standardizing case definition, reporting protocols, and data sharing practices would greatly improve data comparability, transparency, and overall surveillance effectiveness.

Genetic data have proven essential for monitoring viral evolution and reconstructing transmission dynamics through phylogenetic and phylodynamic approaches. However, sequence availability from wild birds remains limited and uneven, introducing significant biases and underrepresenting certain host species and regions. The already complex picture of viral spread[173,174], is further complicated by frequent reassortment events and the co-circulation of multiple genotypes[25,75,175], making it challenging to clearly delineate transmission pathways.

Mortality and survival data are also frequently incomplete or entirely lacking, especially for mass mortality events, resulting in large underestimation of true impacts[176]. Even when such data are available, they rarely provide sufficient detail to assess long-term demographic or species-level consequences. More comprehensive studies are needed to evaluate the ecological effects of outbreaks, particularly on vulnerable species[177–179]. Moreover, critical information is frequently reported in grey literature rather than peer-reviewed sources[180], meaning that literature reviews alone may fail to fully capture the scale of impacts on wild bird populations.

Identifying the key species involved in viral maintenance and dissemination remains particularly challenging. In most studies, infection status and movement data are not integrated[102,108,111,113,126], limiting our ability to determine whether —and to what extent— specific species contribute to transmission[61,81,82]. Integrative approaches that combine genetic, epidemiological, ecological and movement data offer promising avenues to gain deeper insights into species-specific roles and transmission routes[30,47,72,168]. Experimental infection studies can provide valuable complementary insights into host susceptibility and immune responses, although practical and ethical constraints often limit their applicability[166,181–183]. Overcoming these challenges will require strengthened surveillance systems, integrated data infrastructures, and enhanced coordination across disciplines and institutions.

## Methods

### *Spatio-temporal analysis of wild bird cases*

To assess the impacts of HPAI H5 viruses on wild birds in 2020-2023, we conducted a spatiotemporal analysis of reported cases using data from the WAHIS platform[136]. Within the study period, three seasons were defined: 2020-2021, 2021-2022 and 2022-2023, ranging from October 1st to September 30th of the following year (extended to the end of 2023 for the 2022-2023 wave), consistent with the seasonality adopted by EFSA[129]. Due to the limited clade-specific information available in the WAHIS database and the predominance of H5N8 and H5N1 subtypes in reported cases since 2020 (Extended Data Fig. 1)[26], we restricted our analysis on these two subtypes and added H5Nx in case of incomplete subtyping. A case has been defined as a reported case in a wild bird species, per location and per month, as reported to WAHIS. This definition does not account for the quantitative aspects of mortality, notably in mass mortality events. Wild birds were categorized into six groups based on taxonomic, epidemiological, and ecological criteria: Seabirds (excluding Laridae), Laridae, Other waterbirds, Anatidae, Raptors, and Other land birds (Extended Data Table 1).

### *Systematic literature review*

In parallel, we conducted a systematic literature review following the PRISMA-ScR (Preferred Reporting Items for Systematic Reviews and Meta-Analyses Extension for Scoping Reviews) guidelines[184]. The search was conducted in two databases (PubMed and Web of Science) and was last updated on November 2024. The Boolean query combined two groups of keywords ("wild birds" and "HPAI"), linked by the Boolean operator "AND," with keywords within each group connected by the operator "OR" (Extended Data Fig. 2). Searches were restricted to titles and abstracts in all databases. To focus on recent HPAI epizootics, the search included only publications from 2020 onward. Only English written articles published in indexed scientific journals were considered. Articles underwent a two-step screening process for inclusion in the final analysis. The first screening assessed titles and abstracts, while the second evaluated full texts. In both screenings, articles had to meet the following criteria: (1) investigate HPAI H5 viruses, specifically the H5N1 or H5N8 subtypes; (2) focus on events occurring from October 2020 to the end of 2023; (3) analyze the spatial spread of HPAI H5 viruses, assess its impact on wild birds or evaluate infection and transmission pathways. Literature reviews were excluded. Primary and secondary screenings were conducted independently by two authors, using a conservative approach in cases of disagreement. To identify additional relevant articles, we employed a snowball sampling technique to include cited articles not captured in the initial search (Extended Data Fig. 2). Data extraction was performed on the remaining articles, focusing on two key aspects: (1) the spatial spread of HPAI and its links to wild bird movements, (2) infection and transmission pathways in wild birds, as well as the resulting mortality or impacts on their populations.

**Table 1. Major mortality events caused by HPAI H5 viruses in seabirds, Laridae and other waterbirds.** Most events were identified through the literature review but some substantial events were added through cases declared in WAHIS (* denotes that the number of deaths was based on estimations from WAHIS). N_dead represents the estimated number of deaths.

| Groups | Species | IUCN status | Location | Period | Impact on populations | References |
|---|---|---|---|---|---|---|
| **Seabirds** | Northern gannet (*Morus bassanus*) | Least Concern | North Atlantic | Summer 2022 | 75% of colonies affected (41/53) | [107] |
| | | | Scotland | Summer 2022 | <u>Bass Rock:</u><br>Colony 71% smaller compared to last full count,<br>3% of breeding adults dead,<br>75% decline in breeding success<br>N_dead=3,761-5,035 | [107,124,179] |
| | | | | | Alderney:<br>N_dead = 3,500-4,300<br>Faroe islands:<br>7% mortality<br>Sule Skerry:<br>6% mortality | |
| | | | France | Summer 2022 | Rouzic:<br>54% decline of apparently occupied sites<br>58-87% breeding adult mortality<br>67-94% chick mortality | [108] |
| | | | Ireland | Summer 2022 | <4% dead birds<br>N_dead= 1,526 | [185] |
| | | | Germany | Summer 2022 | 8.7% of adult mortality in a colony (N_dead=1,645) | [122] |
| | | | The Netherlands | Summer 2022 | 32.8%–90% of mortality (N_dead =2,215) | [131] |
| | | | Canada (Atlantic coast) | Summer 2022 | 11.5% loss of the North American breeding population (N_dead=25,669)<br><br>Île Bonaventure:<br>3.4% mortality<br><br>Rochers aux Oiseaux:<br>9.9% mortality<br>58% decline in the number of apparently occupied sites<br><br>Cape St-Mary's:<br>17% of reproductive success (lowest recorded since 1970) | [109,111] |
| | | | Russia (Barents Sea) | Summer 2022 | 41.5% decrease in inhabitable nests | [112] |
| | Peruvian booby (*Sula variegata*) | Least Concern | Chile, Peru | Late 2022 | Peru:<br>3.96% of population affected (N_dead=47,531) | [86,89,92,116–118] |
| | Great skua (*Stercorarius skua*) | Least Concern | Scotland | Summer 2021 | >10% of breeding adults dead | [114] |
| | | | Great Britain | Summer 2022 | 11% of the breeding population of Great Britain<br>7% of the world population (N_dead=2,200) | [124] |
| | Brown skua (*Stercorarius antarcticus*) | Least Concern | Antarctic region | Late 2023 | (N_dead=77) | [2,100] |
| | African penguin (*Spheniscus demersus*) | Critically Endangered | South Africa | Late 2021 | N_dead>200 | [103] |
| | | | South Africa | Summer-Late 2022 | N_dead=60 | [103] |
| | Humboldt penguin (*Spheniscus humboldti*) | Vulnerable | Chile | Late 2023 | N_dead=3,000 | [119] |
| | Gentoo penguin | Least Concern | Antarctic region | Late 2023 | N_dead=38 | [100] |
| | (*Pygoscelis papua*) | | | | | |
| | Common murre<br>(*Uria aalge*) | Least Concern | Canada<br>(Atlantic coast) | Summer 2022 | N_dead=8,133<br><0.5% of the breeding population | [109] |
| | Magnificent frigate<br>(*Fregata magnificens*) | Least Concern | Ecuador | Late 2023 | N_dead=6,000 | [120] |
| | Great frigate<br>(*Fregata minor*) | Least Concern | Ecuador | Late 2023 | N_dead=1,000 | [120] |
| | Snowy albatross<br>(*Diomedea exulans*) | Vulnerable | Antarctic region | Late 2023 | N_dead=58 | [100] |
| | Atlantic puffin<br>(*Fratercula arctica*) | Vulnerable | Canada<br>(Atlantic coast) | Summer 2022 | N_dead=282 | [109] |
| | Razorbill<br>(*Alca torda*) | Least Concern | Canada<br>(Atlantic coast) | Summer 2022 | N_dead=119 | [109] |
| **Laridae** | Sandwich tern<br>(*Thalasseus sandvicensis*) | Least Concern | Europe | Summer 2022 | HPAI confirmed in 60% of colonies (39/65)<br>>17% of the total breeding population<br>(N_dead=20,531) | [121,136] |
| | | | Germany | Summer 2022 | >30% adult mortality in 3 colonies<br>(sum of 5 colonies: N_dead>9,497) | [122] |
| | | | Brazil | Summer 2023 | N_dead=343* | [136] |
| | | | The Netherlands | Summer 2022 | >99% mortality in 6 colonies<br>(sum of 10 colonies: N_dead=8,001)<br><br>17.2-90% of mortality | [123,131] |
| | Common tern<br>(*Sterna hirundo*) | Least Concern | Germany | Summer 2022 | >37% adult mortality in 2 colonies<br>(sum of 3 colonies: N_dead=2,316) | [122] |
| | | | Ireland | Summer 2023 | 17% adult mortality in one colony<br>(sum of 3 colonies: N_dead=1,788) | [125] |
| | Caspian tern<br>(*Hydroprogne caspia*) | Least Concern | Russia<br>(Caspian Sea) | Summer 2022 | N_dead=5,641 | [58] |
| | | | United States<br>(Lake Michigan) | Summer 2022 | 62% of mortality<br>(N_dead=1,240) | [19] |
| | | | United States<br>(Pacific coast) | Summer 2023 | <u>Rat island:</u><br>53-56% of adult mortality<br>(N_dead=1,101)<br><br>Pacific :<br>10-14% of adult mortality<br>(N_dead=1,529) | [126] |
| | Arctic tern<br>(*Sterna paradisaea*) | Least Concern | Ireland | Summer 2023 | sum of 3 colonies: N_dead=48 | [125] |
| | Roseate tern<br>(*Sterna dougallii*) | Least Concern | Ireland | Summer 2023 | sum of 3 colonies: N_dead=65 | [125] |
| | Swift tern<br>(*Thalasseus bergii*) | Least Concern | South Africa | Summer 2023 | N_dead=82 | [127] |
| | Royal tern<br>(*Thalasseus maximus*) | Least Concern | Gambia | Early 2023 | N_dead=336* | [136] |
| | South American tern<br>(*Sterna hirundinacea*) | Least Concern | Argentina | Late 2023 | N_dead=400 | [6] |
| | Tern species | | Great Britain | Summer 2022 | hundreds | [124] |
| | Tern species | | Canada (Atlantic coast) | Summer 2022 | N_dead=74 | [109] |
| | Black-headed gull (*Chroicocephalus ridibundus*) | Least Concern | Europe | 2023 | N_dead>5,000* | [26,106,128–130 136] |
| | | | Germany | Summer 2022 | 2.1% adult mortality in one colony (sum of 2 colonies: N_dead=752) | [122] |
| | Great black-backed gull (*Larus marinus*) | Least Concern | The Netherlands | 2020-2021 | N_dead=137 (0.01–5.4% of population) | [131] |
| | | | The Netherlands | 2021-2022 | N_dead=372 (1.4–14.8% of population) | [131] |
| | Black-legged kittiwake (*Rissa tridactyla*) | Vulnerable | Canada (Atlantic coast) | Summer 2022 | N_dead=251 | [109,110] |
| | | | Norway | Summer 2023 | N_dead=12,000* | [136] |
| | Great black-headed gull (*Ichthyaetus ichthyaetus*) | Least Concern | Russia (Caspian Sea) | Summer 2022 | N_dead=25,157 | [58] |
| | Caspian gull (*Larus cachinnans*) | Least Concern | Russia (Caspian Sea) | Summer 2022 | N_dead=3,507 | [58] |
| | Kelp gull (*Larus dominicanus*) | Least Concern | Chile | Late 2022-2023 | N_dead=54 | [116,118] |
| | Gull species | | Canada (Atlantic coast) | Summer 2022 | N_dead=2,326 | [109] |
| | Grey gull (*Leucophaeus modestus*) | Least Concern | Peru | Summer 2023 | N_dead=745* | [136] |
| | Mediterranean gull (*Ichthyaetus melanocephalus*) | Least Concern | France | Summer 2023 | N_dead=309* | [136] |
| | Brown-headed gull (*Chroicocephalus brunnicephalus*) | Least Concern | China | Summer 2022 | N_dead=234* | [136] |
| **Other waterbirds** | Great white pelican (*Pelecanus onocrotalus*) | Least Concern | Senegal | Early 2021 | 8.4% of mortality (N_dead=750) | [44] |
| | | | Mauritania | Early 2021 | N_dead=495* | [136] |
| | | | Israel | Late 2021 | Hundreds | [135] |
| | Dalmatian pelican (*Pelecanus crispus*) | Near Threatened | Greece | Early 2022 | 60% of mortality (N_dead=1,734) | [133,134] |
| | American white pelican (*Pelecanus erythrorhynchos*) | Least Concern | United States | 2021-2023 | N_dead=1,050 N_dead=1,025 | [19] |
| | | | Canada | Summer 2022 | N_dead<100 | [83] |
| | Peruvian pelican (*Pelecanus thagus*) | Near Threatened | Peru, Chile | Late 2022 | 13.4 – 25.3 % of Peruvian population (N_dead= 21,199) 2.11–21.2 % of total population | [23,86,87,89,92,116–118] |
| | Guanay cormorant (*Leucocarbo bougainvilliorum*) | Near Threatened | Peru, Chile | Late 2022 | 0.58–1.16% of total population 1.13–1.39 % of Peruvian population (N_dead= 28,921) | [92,116,118] |
| | Cape cormorant (*Phalacrocorax capensis*) | Endangered | South Africa, Namibia | Late 2021 – Early 2022 | 30% of breeding population (N_dead= 24,000) | [103,137] |
| | Double-crested cormorant (*Nannopterum auritum*) | Least Concern | United States, Canada | 2021-2023 | N_dead=920<br>N_dead=580<br>+ other mass mortalities | [19,83] |
| | Great cormorant (*Phalacrocorax carbo*) | Least Concern | Europe | Summer 2021-2022 | N_dead=1,700 | [132] |
| | Common crane (*Grus grus*) | Least Concern | Israel | Late 2021 | N_dead=10,000 | [135] |
| | Hooded crane (*Grus monacha*) | Vulnerable | Japan, Korea | Late 2022 | N_dead=1,476,<br>N_dead=221<br>10% of population lost | [71] |
| | Black necked grebe (*Podiceps nigricollis*) | Least Concern | China | Summer 2021 | N_dead=4,000 | [138,139] |
| | | | Canada | Summer 2022 | N_dead<100 | [83] |
| | Western grebe (*Aechmophorus occidentalis*) | Least Concern | Canada | Summer 2022 | N_dead<100 | [83] |
| | American coot (*Fulica americana*) | Least Concern | United States | 2021-2023 | N_dead=250,<br>N_dead=160 | [19] |
| | Great egret (*Ardea alba*) | Least Concern | United States | 2021-2023 | N_dead=115 | [19] |
| | James's flamingo (*Phoenicoparrus jamesi*) | Near Threatened | Argentina | Late 2023 | N_dead=220 | [119] |

**Acknowledgments**

We are grateful to Amandine Wanert for creating the schematic overview of the global transmission routes (Fig. 1). We also thank Bernard Cadiou (Bretagne Vivante), Yann Jacob (Bretagne Vivante), Robert Straughan (LPO), Pascal Provost (LPO) and Pierre-Yves Henry (MNHN) for sharing the pictures of affected wild birds (Fig. 2). This work was co-funded by the European Union's Horizon Europe Project 101136346 EUPAHW.

## Extended Data

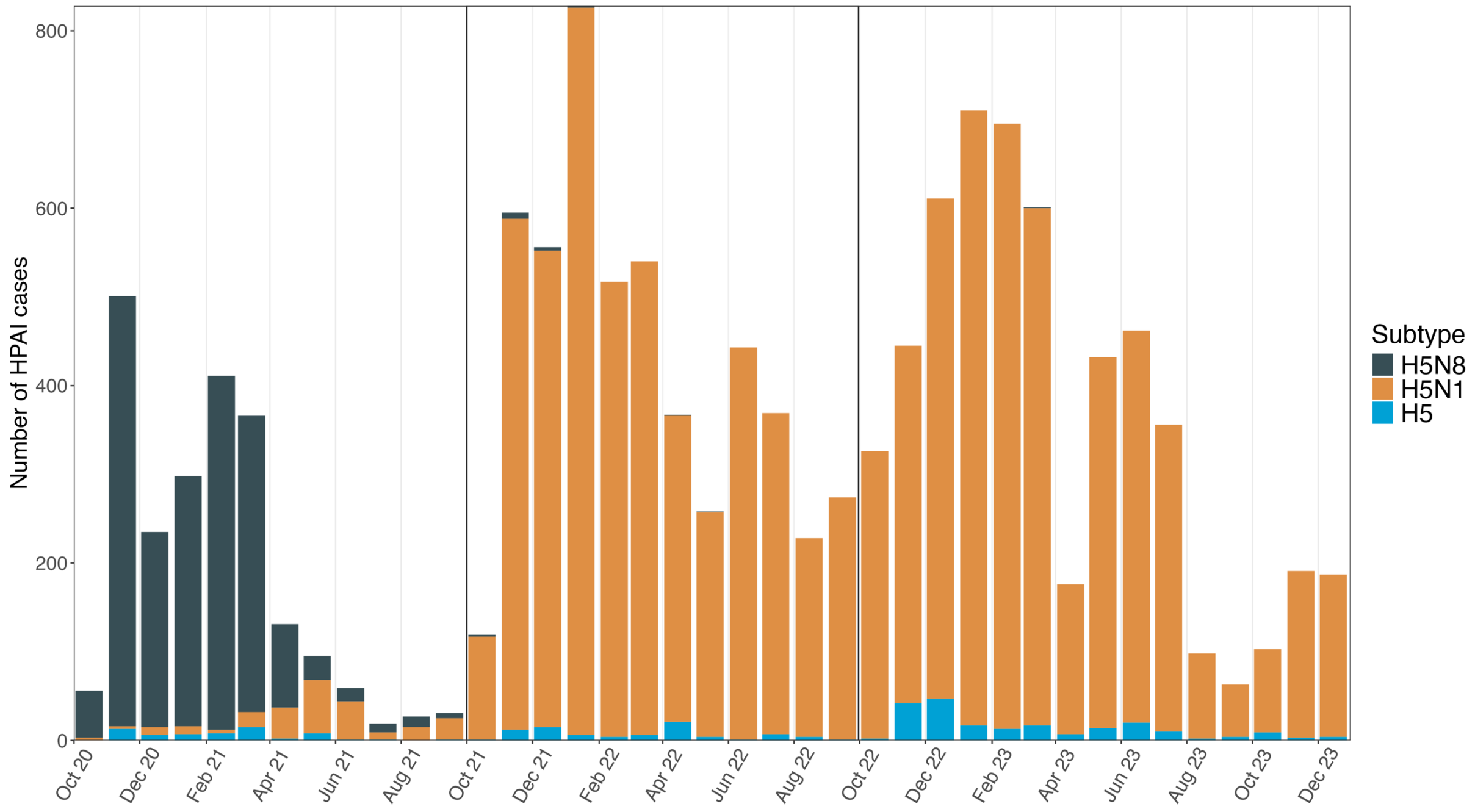

**Extended Data Figure 1:** Temporal distribution of HPAI H5 (H5N8, H5N1, H5Nx) cases in wild birds, as reported to World Animal Health Information System (WAHIS) of the World Organization for Animal Health. A case has been defined as a reported case in a wild bird species, per location and per month, as reported to WAHIS. Seasons were defined from October 1st to September 30$^{th}$ (extended to the end of 2023 for the 2022-2023 wave).

**Extended Data Table 1:** Table indicating which families and orders of wild birds are within each group, as used in this study.

| **Wild bird group** | **Anatidae** | **Laridae** | **Seabirds** | **Other waterbirds** | **Raptors** | **Other land birds** |
|---|---|---|---|---|---|---|
| **Taxonomy** | Anseriformes (Anatidae) | Charadriiformes (Laridae) | Procellariiformes<br>Sphenisciformes<br><br>**Suliformes** (Sulidae, Fregatidae)<br>**Charadriiformes** (Stercorariidae, Alcidae) | Pelecaniformes<br>Gruiformes<br>Ciconiiformes<br>Podicipediformes<br>Phoenicopteriformes<br>Gaviiformes<br>**Suliformes** (Phalacrocoracidae)<br>**Charadriiformes** (Scolopacidae, Charadriidae, Haematopodidae) | Accipitriformes<br>Falconiformes<br>Strigiformes<br>Cathartiformes | Passeriformes<br>Galliformes<br>Columbiformes<br>Psittaciformes<br>Casuariiformes<br>Rheiformes<br>Struthioniformes<br>Piciformes<br>Caprimulgiformes |

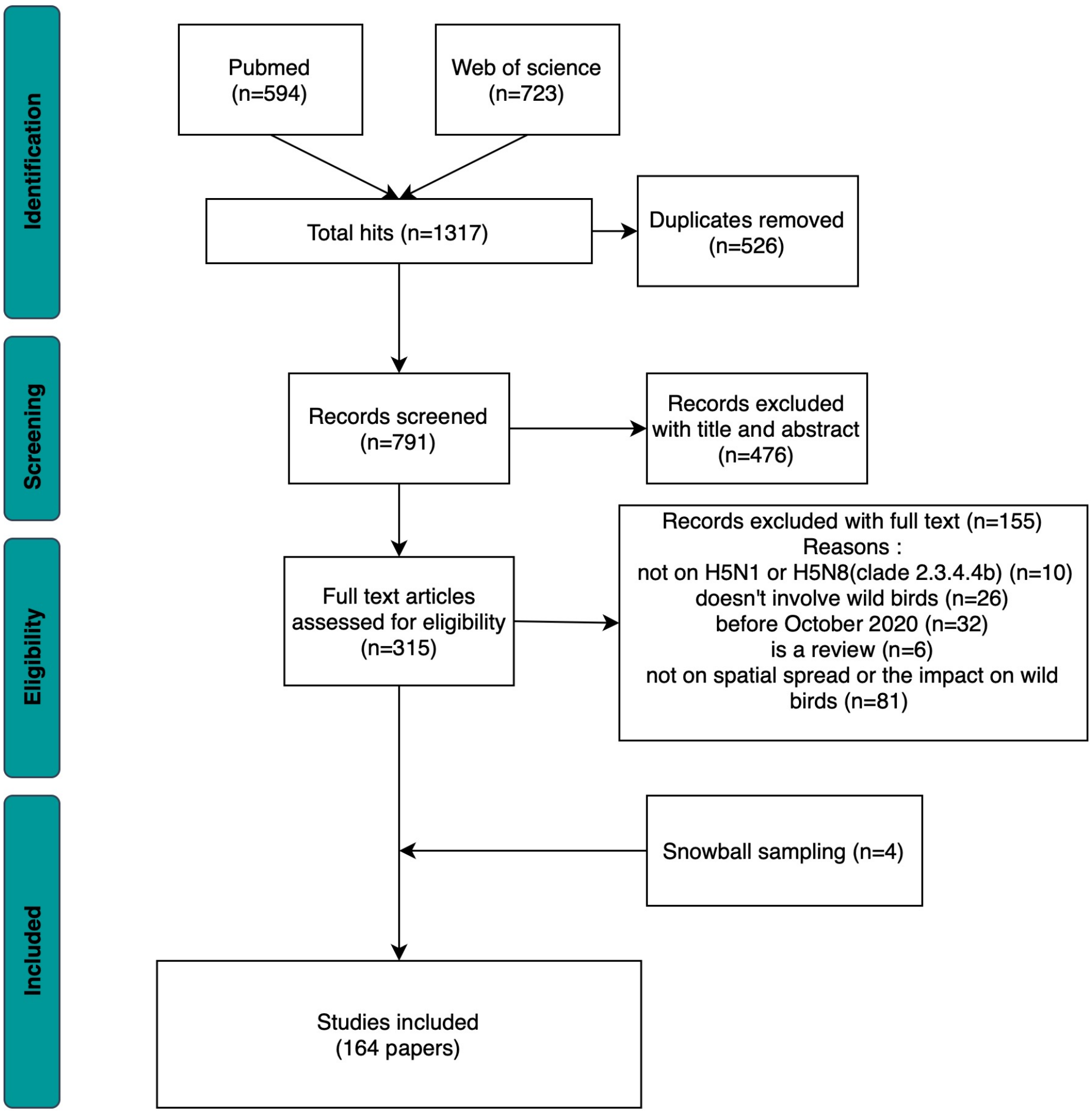

**Extended Data Figure 2:** Prisma-ScR flow diagram of the reviewing process. The boolean query was the following: (“wild bird*” OR “migratory bird*” OR waterfowl* OR seabird* OR Accipitriformes OR Anseriformes OR Charadriiformes OR Laridae OR Anatidae OR tern* OR gull* OR mallard* OR swan* OR raptor* OR falcon* OR gannet* OR murre* OR eagle* OR stork* OR buzzard* OR cormorant*)

AND

(“highly pathogenic avian influenza” OR “highly pathogenic influenza A virus*” OR “highly pathogenic H5*” OR “high pathogeni*” OR HPAI* OR H5N1 OR H5N8 OR H5N*)

**Supplementary data:** All 164 records kept from the systematic review filtering process

| Author | Title | Journal Title | DOI | Date |
|---|---|---|---|---|
| Abolnik, Celia | The Molecular Epidemiology of Clade 2.3.4.4B H5N1 High Pathogenicity Avian Influenza in Southern Africa, 2021-2022 | Viruses-Basel | 10.3390/v15061383 | 2023-06 |
| Abolnik, Celia | Outbreaks of H5N1 High Pathogenicity Avian Influenza in South Africa in 2023 Were Caused by Two Distinct Sub-Genotypes of Clade 2.3.4.4b Viruses | Viruses | 10.3390/v16060896 | 31/05/2024 |
| Ahlstrom, Christina A. | Genomic characterization of highly pathogenic H5 avian influenza viruses from Alaska during 2022 provides evidence for genotype-specific trends of spatiotemporal and interspecies dissemination | Emerging Microbes & Infections | 10.1080/22221751.2024.2406291 | 2024-12 |
| Ahrens, Ann Kathrin | Iceland: an underestimated hub for the spread of high-pathogenicity avian influenza viruses in the North Atlantic | The Journal Of General Virology | 10.1099/jgv.0.001985 | 2024-05 |
| Ahrens, Ann Kathrin | Exploring surface water as a transmission medium of avian influenza viruses – systematic infection studies in mallards | Emerging Microbes & Infections | 10.1080/22221751.2022.2065937 | 31/12/2022 |
| Alexandrou, Olga | The impact of avian influenza 2022 on Dalmatian pelicans was the worst ever wildlife disaster in Greece | Oryx | 10.1017/S0030605322001041 | 2022-11 |
| Alkie, Tamiru N. | Recurring Trans-Atlantic Incursion of Clade 2.3.4.4b H5N1 Viruses by Long Distance Migratory Birds from Northern Europe to Canada in 2022/2023 | Viruses-Basel | 10.3390/v15091836 | 2023-09 |
| Alkie, Tamiru N. | A threat from both sides: Multiple introductions of genetically distinct H5 HPAI viruses into Canada via both East Asia-Australasia/Pacific and Atlantic flyways | Virus Evolution | 10.1093/ve/veac077 | 10/09/2022 |
| Andrew, Cassandra L. | Descriptive Epidemiology and Phylodynamics of the "First Wave" of an Outbreak of Highly Pathogenic Avian Influenza (H5N1 Clade 2.3.4.4b) in British Columbia and the Yukon, Canada, April to September 2022 | Transboundary And Emerging Diseases | 10.1155/2024/2327939 | 29/02/2024 |
| Ariyama, Naomi | Highly Pathogenic Avian Influenza A(H5N1) Clade 2.3.4.4b Virus in Wild Birds, Chile | Emerging Infectious Diseases | 10.3201/eid2909.230067 | 2023-09 |
| Artuso, María Carolina | Detection and characterization of highly pathogenic avian influenza A (H5N1) clade 2.3.4.4b virus circulating in Argentina in 2023 | Revista Argentina De Microbiologia | 10.1016/j.ram.2024.08.002 | 11/11/2024 |
| Avery-Gomm, S | Wild bird mass mortalities in eastern Canada associated with the Highly Pathogenic Avian Influenza A(H5N1) virus, 2022 | Ecosphere | 10.1002/ecs2.4980 | 2024-09 |
| Azat, Claudio | Spatio-temporal dynamics and drivers of highly pathogenic avian influenza H5N1 in Chile | Frontiers In Veterinary Science | 10.3389/fvets.2024.1387040 | 2024 |
| Aznar, Inma | Annual Report on surveillance for avian influenza in poultry and wild birds in Member States of the European Union in 2020 | Efsa Journal | 10.2903/j.efsa.2021.6953 | 2021-12 |
| Aznar, Inma | Annual report on surveillance for avian influenza in poultry and wild birds in Member States of the European Union in 2021 | Efsa Journal | 10.2903/j.efsa.2022.7554 | 2022-09 |
| Banyard, Ashley C. | Detection and spread of high pathogenicity avian influenza virus H5N1 in the Antarctic Region | Nature Communications | 10.1038/s41467-024-51490-8 | 03/09/2024 |
| Banyard, Ashley C. | Detection of Highly Pathogenic Avian Influenza Virus H5N1 Clade 2.3.4.4b in Great Skuas: A Species of Conservation Concern in Great Britain | Viruses-Basel | 10.3390/v14020212 | 2022-02 |
| Bennison, A | A case study of highly pathogenic avian influenza (HPAI) H5N1 at Bird Island, South Georgia: the first documented outbreak in the subantarctic region | Bird Study | 10.1080/00063657.2024.2396563 | 27/09/2024 |
| Bevins, Sarah N. | Intercontinental Movement of Highly Pathogenic Avian Influenza A(H5N1) Clade 2.3.4.4 Virus to the United States, 2021 | Emerging Infectious Diseases | 10.3201/eid2805.220318 | 2022-05 |
| Bøe, Cathrine Arnason | Emergence of highly pathogenic avian influenza viruses H5N1 and H5N5 in white-tailed eagles, 2021-2023 | The Journal Of General Virology | 10.1099/jgv.0.002035 | 2024-11 |
| Bordes, Luca | Experimental infection of chickens, Pekin ducks, Eurasian wigeons and Barnacle geese with two recent highly pathogenic avian influenza H5N1 clade 2.3.4.4b viruses | Emerging Microbes & Infections | 10.1080/22221751.2024.2399970 | 2024-12 |
| Boulinier, Thierry | Avian influenza spread and seabird movements between colonies | Trends In Ecology & Evolution | 10.1016/j.tree.2023.02.002 | 2023-05 |
| Bregnballe, T | Outbreaks of highly pathogenic avian influenza (HPAI) epidemics in Baltic Great Cormorant Phalacrocorax carbo colonies in 2021 and 2022 | Bird Study | 10.1080/00063657.2024.2399168 | 27/09/2024 |
| Briand, Francois-Xavier | Multiple independent introductions of highly pathogenic avian influenza H5 viruses during the 2020-2021 epizootic in France | Transboundary And Emerging Diseases | 10.1111/tbed.14711 | 2022-11 |
| Bruno, A | Highly Pathogenic Avian Influenza A (H5N1) Virus Outbreak in Ecuador in 2022-2024 | Current Infectious Disease Reports | 10.1007/s11908-024-00849-5 | 2024-12 |
| Burke, B | A case study of the 2023 highly pathogenic avian influenza (HPAI) outbreak in tern (Sternidae) colonies on the east coast of the Republic of Ireland | Bird Study | 10.1080/00063657.2024.2409196 | 30/10/2024 |
| Caliendo, V | Transatlantic spread of highly pathogenic avian influenza H5N1 by wild birds from Europe to North America in 2021 | Scientific Reports | 10.1038/s41598-022-13447-z | 11/07/2022 |
| Caliendo, Valentina | Effect of 2020-21 and 2021-22 Highly Pathogenic Avian Influenza H5 Epidemics on Wild Birds, the Netherlands | Emerging Infectious Diseases | 10.3201/eid3001.230970 | 2024-01 |
| Caliendo, Valentina | Pathology and virology of natural highly pathogenic avian influenza H5N8 infection in wild Common buzzards (<i>Buteo buteo</i>) | Scientific Reports | 10.1038/s41598-022-04896-7 | 18/01/2022 |
| Careen, NG | Highly pathogenic avian influenza resulted in unprecedented reproductive failure and movement behaviour by Northern Gannets | Marine Ornithology | 10.5038/2074-1235.52.1.1566 | 2024-04 |
| Castro-Sanguinetti, Gina R. | Highly pathogenic avian influenza virus H5N1 clade 2.3.4.4b from Peru forms a monophyletic group with Chilean isolates in South America | Scientific Reports | 10.1038/s41598-024-54072-2 | 13/02/2024 |
| Cha, Ra Mi | Genetic Characterization and Pathogenesis of H5N1 High Pathogenicity Avian Influenza Virus Isolated in South Korea during 2021-2022 | Viruses-Basel | 10.3390/v15061403 | 2023-06 |
| Cho, Andrew Yong | Index case of H5N1 clade 2.3.4.4b highly pathogenic avian influenza virus in wild birds, South Korea, November 2023 | Frontiers In Veterinary Science | 10.3389/fvets.2024.1366082 | 2024 |
| Cormier, TL | Seabird and sea duck mortalities were lower during the second breeding season in eastern Canada following the introduction of highly pathogenic avian influenza A H5Nx viruses | Bird Study | 10.1080/00063657.2024.2415161 | 30/10/2024 |
| Cruz, Cristopher D. | Highly Pathogenic Avian Influenza A(H5N1) from Wild Birds, Poultry, and Mammals, Peru | Emerging Infectious Diseases | 10.3201/eid2912.230505 | 2023-12 |
| de Araújo, Andreina Carvalho | Incursion of Highly Pathogenic Avian Influenza A(H5N1) Clade 2.3.4.4b Virus, Brazil, 2023 | Emerging Infectious Diseases | 10.3201/eid3003.231157 | 2024-03 |
| Duff, Paul | Investigations associated with the 2020/21 highly pathogenic avian influenza epizootic in wild birds in Great Britain | The Veterinary Record | 10.1002/vetr.1146 | 2021-11 |
| Duriez, Olivier | Highly pathogenic avian influenza affects vultures' movements and breeding output | Current Biology: Cb | 10.1016/j.cub.2023.07.061 | 11/09/2023 |
| Dziadek, K | Phylogenetic and Molecular Characteristics of Wild Bird-Origin Avian Influenza Viruses Circulating in Poland in 2018-2022: Reassortment, Multiple Introductions, and Wild Bird-Poultry Epidemiological Links | Transboundary And Emerging Diseases | 10.1155/2024/6661672 | 12/04/2024 |
| EFSA | Avian influenza overview December 2022-March 2023 | Efsa Journal | 10.2903/j.efsa.2023.7917 | 2023-03 |
| EFSA | Avian influenza overview - update on 19 November 2020, EU/EEA and the UK | Efsa Journal | 10.2903/j.efsa.2020.6341 | 2020-11 |
| EFSA | Avian influenza overview December 2021-March 2022 | Efsa Journal | 10.2903/j.efsa.2022.7289 | 2022-04 |
| EFSA | Avian influenza overview May - September 2021 | Efsa Journal | 10.2903/j.efsa.2022.7122 | 2022-01 |
| EFSA | Avian influenza overview September - December 2021 | Efsa Journal | 10.2903/j.efsa.2021.7108 | 2021-12 |
| EFSA | Avian influenza overview March - June 2022 | Efsa Journal | 10.2903/j.efsa.2022.7415 | 2022-08 |
| EFSA | Avian influenza overview September - December 2022 | Efsa Journal | 10.2903/j.efsa.2023.7786 | 2023-01 |
| EFSA | Avian influenza overview February - May 2021 | Efsa Journal | 10.2903/j.efsa.2021.6951 | 2021-12 |
| EFSA | Avian influenza overview August - December 2020 | Efsa Journal | 10.2903/j.efsa.2020.6379 | 2020-12 |
| EFSA | Avian influenza overview June - September 2022 | Efsa Journal | 10.2903/j.efsa.2022.7597 | 2022-10 |
| EFSA | Avian influenza overview December 2020 - February 2021 | Efsa Journal | 10.2903/j.efsa.2021.6497 | 2021-03 |
| EFSA | Avian influenza overview April - June 2023 | Efsa Journal | 10.2903/j.efsa.2023.8191 | 2023-07 |
| EFSA | Avian influenza overview March - April 2023 | Efsa Journal | 10.2903/j.efsa.2023.8039 | 2023-06 |
| EFSA | Avian influenza overview June-September 2023 | Efsa Journal | 10.2903/j.efsa.2023.8328 | 2023-10 |
| EFSA | Annual report on surveillance for avian influenza in poultry and wild birds in Member States of the European Union in 2022 | Efsa Journal | 10.2903/j.efsa.2023.8480 | 2023-12 |
| El-Sheshenv, Rabeh | Highly Pathogenic Avian Influenza A(H5N1) Virus Clade 2.3.4.4b in Wild Birds and Live Bird Markets, Egypt | Pathogens | 10.3390/pathogens12010036 | 2023-01 |
| Engelsma, Marc | Multiple Introductions of Reassorted Highly Pathogenic Avian Influenza H5Nx Viruses Clade 2.3.4.4b Causing Outbreaks in Wild Birds and Poultry in The Netherlands, 2020-2021 | Microbiology Spectrum | 10.1128/spectrum.02499-21 | 2022-04 |
| European Food Safety Authority | Avian influenza overview September-December 2023 | Efsa Journal. European Food Safety Authority | 10.2903/j.efsa.2023.8539 | 2023-12 |
| Falchieri, Marco | Shift in HPAI infection dynamics causes significant losses in seabird populations across Great Britain | Veterinary Record | 10.1002/vetr.2311 | 2022-10 |
| Fujimoto, Yoshikazu | Experimental and natural infections of white-tailed sea eagles (<i>Haliaeetus albicilla</i>) with high pathogenicity avian influenza virus of H5 subtype | Frontiers In Microbiology | 10.3389/fmicb.2022.1007350 | 03/10/2022 |
| Fujita, Ryosuke | Blowflies are potential vector for avian influenza virus at enzootic area in Japan | Scientific Reports | 10.1038/s41598-024-61026-1 | 04/05/2024 |
| Furness, Robert W. | Environmental Samples Test Negative for Avian Influenza Virus H5N1 Four Months after Mass Mortality at A Seabird Colony | Pathogens | 10.3390/pathogens12040584 | 2023-04 |
| Fusaro, Alice | High pathogenic avian influenza A(H5) viruses of clade 2.3.4.4b in Europe-Why trends of virus evolution are more difficult to predict | Virus Evolution | 10.1093/ve/veae027 | 2024 |
| Gamarra-Toledo, Victor | Highly Pathogenic Avian Influenza (HPAI) strongly impacts wild birds in Peru | Biological Conservation | 10.1016/j.biocon.2023.110272 | 2023-10 |
| Gamarra-Toledo, Víctor | Avian flu threatens Neotropical birds | Science | 10.1126/science.adg2271 | 20/01/2023 |
| Gass, Jonathon D. | Ecogeographic Drivers of the Spatial Spread of Highly Pathogenic Avian Influenza Outbreaks in Europe and the United States, 2016-Early 2022 | International Journal Of Environmental Research And Public Health | 10.3390/ijerph20116030 | 01/06/2023 |
| Giacinti, Jolene A. | Avian influenza viruses in wild birds in Canada following incursions of highly pathogenic H5N1 virus from Eurasia in 2021-2022 | Mbio | 10.1128/mbio.03203-23 | 14/08/2024 |
| Glazunova, Anastasia | A highly pathogenic avian influenza virus H5N1 clade 2.3.4.4 detected in Samara Oblast, Russian Federation | Frontiers In Veterinary Science | 10.3389/fvets.2024.1244430 | 2024 |
| Gobbo, Federica | Active Surveillance for Highly Pathogenic Avian Influenza Viruses in Wintering Waterbirds in Northeast Italy, 2020-2021 | Microorganisms | 10.3390/microorganisms9112188 | 2021-11 |
| Godoy, Marcos | Evolution and Current Status of Influenza A Virus in Chile: A Review | Pathogens (Basel, Switzerland) | 10.3390/pathogens12101252 | 17/10/2023 |
| Gorta, SBZ | Kleptoparasitism in seabirds-A potential pathway for global avian influenza virus spread | Conservation Letters | 10.1111/conl.13052 | 16/09/2024 |
| Grant, Malin | Highly Pathogenic Avian Influenza (HPAI H5Nx, Clade 2.3.4.4.b) in Poultry and Wild Birds in Sweden: Synopsis of the 2020-2021 Season | Veterinary Sciences | 10.3390/vetsci9070344 | 2022-07 |
| Grémillet, David | Strong breeding colony fidelity in northern gannets following high pathogenicity avian influenza virus (HPAIV) outbreak | Biological Conservation | 10.1016/j.biocon.2023.110269 | 01/10/2023 |
| Günther, Anne | Avian raptors are indicator species and victims of high pathogenicity avian influenza virus HPAIV H5N1 (clade 2.3.4.4b) in Germany | Scientific Reports | 10.1038/s41598-024-79930-x | 20/11/2024 |
| Günther, Anne | Iceland as Stepping Stone for Spread of Highly Pathogenic Avian Influenza Virus between Europe and North America | Emerging Infectious Diseases | 10.3201/eid2812.221086 | 2022-12 |
| Günther, Anne | Continuous surveillance of potentially zoonotic avian pathogens detects contemporaneous occurrence of highly pathogenic avian influenza viruses (HPAIV H5) and flaviviruses (USUV, WNV) in several wild and captive birds | Emerging Microbes & Infections | 10.1080/22221751.2023.2231561 | 08/12/2023 |
| Hall, Victoria | Surveillance for highly pathogenic avian influenza A (H5N1) in a raptor rehabilitation center-2022 | Plos One | 10.1371/journal.pone.0299330 | 2024 |
| Haman, Katherine H. | A comprehensive epidemiological approach documenting an outbreak of H5N1 highly pathogenic avian influenza virus clade 2.3.4.4b among gulls, terns, and harbor seals in the Northeastern Pacific | Frontiers In Veterinary Science | 10.3389/fvets.2024.1483922 | 2024 |
| Harvey, Johanna A. | The changing dynamics of highly pathogenic avian influenza H5N1: Next steps for management & science in North America | Biological Conservation | 10.1016/j.biocon.2023.110041 | 2023-06 |
| He, Guimei | Genetically Divergent Highly Pathogenic Avian Influenza A(H5N8) Viruses in Wild Birds, Eastern China | Emerging Infectious Diseases | 10.3201/eid2711.204893 | 2021-11 |
| He, Zhen | Emerging Highly Pathogenic Avian Influenza (H5N8) Virus in <i>Podiceps nigricollis</i> in Northwest China in 2021 | Transboundary And Emerging Diseases | 10.1155/2023/7896376 | 21/02/2023 |
| Hew, LY | Continuous Introduction of H5 High Pathogenicity Avian Influenza Viruses in Hokkaido, Japan: Characterization of Viruses Isolated in Winter 2022-2023 and Early Winter 2023-2024 | Transboundary And Emerging Diseases | 10.1155/2024/1199876 | 14/03/2024 |
| Im, C | Mapping African Swine Fever and Highly Pathogenic Avian Influenza Outbreaks along the Demilitarized Zone in the Korean Peninsula | Transboundary And Emerging Diseases | 10.1155/2024/8824971 | 30/05/2024 |
| Isoda, Norikazu | Detection of New H5N1 High Pathogenicity Avian Influenza Viruses in Winter 2021-2022 in the Far East, Which Are Genetically Close to Those in Europe | Viruses-Basel | 10.3390/v14102168 | 2022-10 |
| Isoda, Norikazu | Re-Invasion of H5N8 High Pathogenicity Avian Influenza Virus Clade 2.3.4.4b in Hokkaido, Japan, 2020 | Viruses-Basel | 10.3390/v12121439 | 2020-12 |
| Jeglinski, Jana W. E. | HPAIV outbreak triggers short-term colony connectivity in a seabird metapopulation | Scientific Reports | 10.1038/s41598-024-53550-x | 07/02/2024 |
| Jimenez-Bluhm, Pedro | Detection and phylogenetic analysis of highly pathogenic A/H5N1 avian influenza clade 2.3.4.4b virus in Chile, 2022 | Emerging Microbes & Infections | 10.1080/22221751.2023.2220569 | 08/12/2023 |
| Kang, Yong-Myung | Introduction of Multiple Novel High Pathogenicity Avian Influenza (H5N1) Virus of Clade 2.3.4.4b into South Korea in 2022 | Transboundary And Emerging Diseases | 10.1155/2023/8339427 | 13/04/2023 |
| Kim, Ji-Yun | Genomic epidemiology of highly pathogenic avian influenza A (H5N1) virus in wild birds in South Korea during 2021-2022: Changes in viral epidemic patterns | Virus Evolution | 10.1093/ve/veae014 | 2024 |
| King, Jacqueline | Highly pathogenic avian influenza virus incursions of subtype H5N8, H5N5, H5N1, H5N4, and H5N3 in Germany during 2020-21 | Virus Evolution | 10.1093/ve/veac035 | 25/04/2022 |
| Knief, Ulrich | Highly pathogenic avian influenza causes mass mortality in Sandwich Tern Thalasseus sandvicensis breeding colonies across north-western Europe | Bird Conservation International | 10.1017/S0959270923000400 | 2024-01 |
| Lane, Jude V. | High pathogenicity avian influenza (H5N1) in Northern Gannets (Morus bassanus): Global spread, clinical signs and demographic consequences | Ibis | 10.1111/ibi.13275 | 2023 |
| Leguia, Mariana | Highly pathogenic avian influenza A (H5N1) in marine mammals and seabirds in Peru | Nature Communications | 10.1038/s41467-023-41182-0 | 07/09/2023 |
| Letsholo, Samantha L. | Emergence of High Pathogenicity Avian Influenza Virus H5N1 Clade 2.3.4.4b in Wild Birds and Poultry in Botswana | Viruses-Basel | 10.3390/v14122601 | 2022-12 |
| Lewis, Nicola S. | Emergence and spread of novel H5N8, H5N5 and H5N1 clade 2.3.4.4 highly pathogenic avian influenza in 2020 | Emerging Microbes & Infections | 10.1080/22221751.2021.1872355 | 01/01/2021 |
| Li, MH | Spatiotemporal and Species-Crossing Transmission Dynamics of Subclade 2.3.4.4b H5Nx HPAIVs | Transboundary And Emerging Diseases | 10.1155/2024/2862053 | 10/07/2024 |
| Li, Xiang | Emergence, prevalence, and evolution of H5N8 avian influenza viruses in central China, 2020 | Emerging Microbes & Infections | 10.1080/22221751.2021.2011622 | 31/12/2022 |
| Liang, Yuan | Novel Clade 2.3.4.4b Highly Pathogenic Avian Influenza A H5N8 and H5N5 Viruses in Denmark, 2020 | Viruses-Basel | 10.3390/v13050886 | 2021-05 |
| Lisovski, Simeon | Unexpected Delayed Incursion of Highly Pathogenic Avian Influenza H5N1 (Clade 2.3.4.4b) Into the Antarctic Region | Influenza And Other Respiratory Viruses | 10.1111/irv.70010 | 2024-10 |
| Liu, Yangfan | Risk factors for avian influenza in Danish poultry and wild birds during the epidemic from June 2020 to May 2021 | Frontiers In Veterinary Science | 10.3389/fvets.2024.1358995 | 21/02/2024 |
| Lo, Fatou T. | Intercontinental Spread of Eurasian Highly Pathogenic Avian Influenza A(H5N1) to Senegal | Emerging Infectious Diseases | 10.3201/eid2801.211401 | 2022-01 |
| Lublin, A. | The History of Highly-Pathogenic Avian Influenza in Israel (H5-subtypes): from 2006 to 2023 | Israel Journal Of Veterinary Medicine | | 2023-06 |
| Lv, Xinru | Highly Pathogenic Avian Influenza A(H5N8) Clade 2.3.4.4b Viruses in Satellite-Tracked Wild Ducks, Ningxia, China, 2020 | Emerging Infectious Diseases | 10.3201/eid2805.211580 | 2022-05 |
| Malmberg, Jennifer L. | Mortality in Wild Turkeys (Meleagris gallopavo) Associated with Natural Infection with H5N1 Highly Pathogenic Avian Influenza Virus (HPAIV) Subclade 2.3.4.4 | Journal Of Wildlife Diseases | 10.7589/JWD-D-22-00161 | 01/10/2023 |
| Marandino, Ana | Spreading of the High-Pathogenicity Avian Influenza (H5N1) Virus of Clade 2.3.4.4b into Uruguay | Viruses-Basel | 10.3390/v15091906 | 2023-09 |
| McDuie, F | Mitigating Risk: Predicting H5N1 Avian Influenza Spread with an Empirical Model of Bird Movement | Transboundary And Emerging Diseases | 10.1155/2024/5525298 | 18/07/2024 |
| McLaughlin, A | Spatiotemporal patterns of low and highly pathogenic avian influenza virus prevalence in murres in Canada from 2007 to 2022--a case study for wildlife viral monitoring | Facets | 10.1139/facets-2023-0185 | 10/07/2024 |
| Mihiretu, Berihun Dires | Novel Genotype of HA Clade 2.3.4.4b H5N8 Subtype High Pathogenicity Avian Influenza Virus Emerged at a Wintering Site of Migratory Birds in Japan, 2021/22 Winter | Pathogens (Basel, Switzerland) | 10.3390/pathogens13050380 | 02/05/2024 |
| Mine, Junki | Genetics of H5N1 and H5N8 High-Pathogenicity Avian Influenza Viruses Isolated in Japan in Winter 2021-2022 | Viruses | 10.3390/v16030358 | 26/02/2024 |
| Mine, Junki | Genetics of Japanese H5N8 high pathogenicity avian influenza viruses isolated in winter 2020-2021 and their genetic relationship with avian influenza viruses in Siberia | Transboundary And Emerging Diseases | 10.1111/tbed.14559 | 2022-09 |
| Molini, Umberto | Highly pathogenic avian influenza H5N1 virus outbreak among Cape cormorants (<i>Phalacrocorax capensis</i>) in Namibia, 2022 | Emerging Microbes & Infections | 10.1080/22221751.2023.2167610 | 31/12/2023 |
| Mosaad, Zienab | Emergence of Highly Pathogenic Avian Influenza A Virus (H5N1) of Clade 2.3.4.4b in Egypt, 2021-2022 | Pathogens | 10.3390/pathogens12010090 | 2023-01 |
| Nagy, Alexander | Enzootic Circulation, Massive Gull Mortality and Poultry Outbreaks during the 2022/2023 High-Pathogenicity Avian Influenza H5N1 Season in the Czech Republic | Viruses | 10.3390/v16020221 | 31/01/2024 |
| Nemeth, Nicole M. | Bald eagle mortality and nest failure due to clade 2.3.4.4 highly pathogenic H5N1 influenza a virus | Scientific Reports | 10.1038/s41598-023-27446-1 | 05/01/2023 |
| Okuya, Kosuke | Isolation and genetic characterization of multiple genotypes of both H5 and H7 avian influenza viruses from environmental water in the Izumi plain, Kagoshima prefecture, Japan during the 2021/22 winter season | Comparative Immunology, Microbiology And Infectious Diseases | 10.1016/j.cimid.2024.102182 | 2024-06 |
| Okuya, Kosuke | Newly emerged genotypes of highly pathogenic H5N8 avian influenza viruses in Kagoshima prefecture, Japan during winter 2020/21 | Journal Of General Virology | 10.1099/jgv.0.001870 | 2023 |
| Olawuyi, Kayode | Detection of clade 2.3.4.4 highly pathogenic avian influenza H5 viruses in healthy wild birds in the Hadeji-Nguru wetland, Nigeria 2022 | Influenza And Other Respiratory Viruses | 10.1111/irv.13254 | 2024-02 |
| Ospina-Jimenez, Andres F. | Sequence-based epitope mapping of high pathogenicity avian influenza H5 clade 2.3.4.4b in Latin America | Frontiers In Veterinary Science | 10.3389/fvets.2024.1347509 | 2024 |
| Paradell, Oriol Giralt | Estimated mortality of the highly pathogenic avian influenza pandemic on northern gannets (<i>Morus bassanus</i>) in southwest Ireland | Biology Letters | 10.1098/rsbl.2023.0090 | 14/06/2023 |
| Pardo-Roa, Catalina | Cross-species transmission and PB2 mammalian adaptations of highly pathogenic avian influenza A/H5N1 viruses in Chile | Biorxiv: The Preprint Server For Biology | 10.1101/2023.06.30.547205 | 30/06/2023 |
| Paternina, Daniela | Dramatic re-emergence of avian influenza in Colombia and Latin America | Travel Medicine And Infectious Disease | 10.1016/j.tmaid.2024.102711 | 2024 |
| Pohlmann, Anne | Mass mortality among colony- breeding seabirds in the German Wadden Sea in 2022 due to distinct genotypes of HPAIV H5N1 clade 2.3.4.4b | Journal Of General Virology | 10.1099/jgv.0.001834 | 2023 |
| Prosser, Diann J. | A lesser scaup (Aythya affinis) naturally infected with Eurasian 2.3.4.4 highly pathogenic H5N1 avian influenza virus: Movement ecology and host factors | Transboundary And Emerging Diseases | 10.1111/tbed.14614 | 2022 |
| Provencher, Jennifer F. | Pathogen Surveillance in Swallows (family Hirundinidae): Investigation into Role as Avian Influenza Vector in Eastern Canada Agricultural Landscapes | Journal Of Wildlife Diseases | 10.7589/JWD-D-23-00167 | 01/07/2024 |
| Puryear, Wendy | Highly Pathogenic Avian Influenza A(H5N1) Virus Outbreak in New England Seals, United States | Emerging Infectious Diseases | 10.3201/eid2904.221538 | 2023-04 |
| Ramey, Andrew M. | Molecular detection and characterization of highly pathogenic H5N1 clade 2.3.4.4b avian influenza viruses among hunter-harvested wild birds provides evidence for three independent introductions into Alaska | Virology | 10.1016/j.virol.2023.109938 | 2024-01 |
| Reid, Scott M. | A multi-species, multi-pathogen avian viral disease outbreak event: Investigating potential for virus transmission at the wild bird - poultry interface | Emerging Microbes & Infections | 10.1080/22221751.2024.2348521 | 2024-12 |
| Reischak, Dilmara | First report and genetic characterization of the highly pathogenic avian influenza A(H5N1) virus in Cabot's tern (Thalasseus acuflavidus), Brazil | Veterinary And Animal Science | 10.1016/j.vas.2023.100319 | 2023-12 |
| Riaz, J | Coastal connectivity of marine predators over the Patagonian Shelf during the highly pathogenic avian influenza outbreak | Ecography | 10.1111/ecog.07415 | 2024-11 |
| Rijks, Jolianne M. | Mass Mortality Caused by Highly Pathogenic Influenza A(H5N1) Virus in Sandwich Terns, the Netherlands, 2022 | Emerging Infectious Diseases | 10.3201/eid2812.221292 | 2022-12 |
| Rimondi, Agustina | Highly Pathogenic Avian Influenza A(H5N1) Viruses from Multispecies Outbreak, Argentina, August 2023 | Emerging Infectious Diseases | 10.3201/eid3004.231725 | 2024-04 |
| Ringenberg, JM | Prevalence of Avian Influenza Virus in Atypical Wild Birds Host Groups during an Outbreak of Highly Pathogenic Strain EA/AM H5N1 | Transboundary And Emerging Diseases | 10.1155/2024/4009552 | 29/07/2024 |
| Rivetti, Anselmo Vasconcelos | Phylodynamics of avian influenza A(H5N1) viruses from outbreaks in Brazil | Virus Research | 10.1016/j.virusres.2024.199415 | 2024-09 |
| Ross, Craig S. | Genetic Analysis of H5N1 High-Pathogenicity Avian Influenza Virus following a Mass Mortality Event in Wild Geese on the Solway Firth | Pathogens (Basel, Switzerland) | 10.3390/pathogens13010083 | 17/01/2024 |
| Russell, Shannon L. | Descriptive epidemiology and phylogenetic analysis of highly pathogenic avian influenza H5N1 clade 2.3.4.4b in British Columbia (B.C.) and the Yukon, Canada, September 2022 to June 2023 | Emerging Microbes & Infections | 10.1080/22221751.2024.2392667 | 2024-12 |
| Sagong, Mingeun | Emergence of clade 2.3.4.4b novel reassortant H5N1 high pathogenicity avian influenza virus in South Korea during late 2021 | Transboundary And Emerging Diseases | 10.1111/tbed.14551 | 2022-09 |
| Seekings, Amanda H. | Transmission dynamics and pathogenesis differ between pheasants and partridges infected with clade 2.3.4.4b H5N8 and H5N1 high- pathogenicity avian influenza viruses | Journal Of General Virology | 10.1099/jgv.0.001946 | 2024 |
| Seo, Ye-Ram | Evolution and Spread of Highly Pathogenic Avian Influenza A(H5N1) Clade 2.3.4.4b Virus in Wild Birds, South Korea, 2022-2023 | Emerging Infectious Diseases | 10.3201/eid3002.231274 | 2024-02 |
| Seo, Ye-Ram | Genetic and pathological analysis of hooded cranes (Grus monacha) naturally infected with clade 2.3.4.4b highly pathogenic avian influenza H5N1 virus in South Korea in the winter of 2022 | Frontiers In Veterinary Science | 10.3389/fvets.2024.1499440 | 2024 |
| Shemmings-Payne, Wesley | Repeatability and reproducibility of hunter-harvest sampling for avian influenza virus surveillance in Great Britain | Research In Veterinary Science | 10.1016/j.rvsc.2024.105279 | 2024-06 |
| Si, Young Jae | Evolutional dynamics of highly pathogenic avian influenza H5N8 genotypes in wintering bird habitats: Insights from South Korea's 2020-2021 season | One Health (Amsterdam, Netherlands) | 10.1016/j.onehlt.2024.100719 | 2024-06 |
| Smietanka, Krzysztof | Highly pathogenic avian influenza H5Nx in Poland in 2020/2021: a descriptive epidemiological study of a large-scale epidemic | Journal Of Veterinary Research | 10.2478/jvetres-2022-0017 | 25/03/2022 |
| Sobolev, Ivan | Highly Pathogenic Avian Influenza A(H5N1) Virus-Induced Mass Death of Wild Birds, Caspian Sea, Russia, 2022 | Emerging Infectious Diseases | 10.3201/eid2912.230330 | 2023-12 |
| Tammiranta, Niina | Highly pathogenic avian influenza A (H5N1) virus infections in wild carnivores connected to mass mortalities of pheasants in Finland | Infection, Genetics And Evolution | 10.1016/j.meegid.2023.105423 | 2023-07 |
| Taylor, Liam U. | Limited Outbreak of Highly Pathogenic Influenza A(H5N1) in Herring Gull Colony, Canada, 2022 | Emerging Infectious Diseases | 10.3201/eid2910.230536 | 2023-10 |
| Teitelbaum, Claire S. | North American wintering mallards infected with highly pathogenic avian influenza show few signs of altered local or migratory movements | Scientific Reports | 10.1038/s41598-023-40921-z | 02/09/2023 |
| Tian, Jingman | Highly Pathogenic Avian Influenza Virus (H5N1) Clade 2.3.4.4b Introduced by Wild Birds, China, 2021 | Emerging Infectious Diseases | 10.3201/eid2907.221149 | 2023-07 |
| Tomás, Gonzalo | Highly pathogenic avian influenza H5N1 virus infections in pinnipeds and seabirds in Uruguay: Implications for bird-mammal transmission in South America | Virus Evolution | 10.1093/ve/veae031 | 2024 |
| Tyndall, Amy A. | Quantifying the Impact of Avian Influenza on the Northern Gannet Colony of Bass Rock Using Ultra-High-Resolution Drone Imagery and Deep Learning | Drones | 10.3390/drones8020040 | 2024-02 |
| Uhart, Marcela M. | Epidemiological data of an influenza A/H5N1 outbreak in elephant seals in Argentina indicates mammal-to-mammal transmission | Nature Communications | 10.1038/s41467-024-53766-5 | 11/11/2024 |
| V. Ezhov, A. | Current State and Development Factors of Northern Gannet (Morus bassanus, Sulidae, Aves) Colonies in the Russian Sector of the Barents Sea | Zoologichesky Zhurnal | 10.31857/S0044513424010076 | 2024-01 |
| Verma, Asha Kumari | Highly pathogenic avian influenza (H5N1) infection in crows through ingestion of infected crow carcasses | Microbial Pathogenesis | 10.1016/j.micpath.2023.106330 | 2023-10 |
| Xie, Ruopeng | The episodic resurgence of highly pathogenic avian influenza H5 virus | Nature | 10.1038/s41586-023-06631-2 | 2023-10 |
| Xiong, Jiasong | Emerging highly pathogenic avian influenza (H5N8) virus in migratory birds in Central China, 2020 | Emerging Microbes & Infections | 10.1080/22221751.2021.1956372 | 01/01/2021 |
| Xu, Qiuzi | Emergence of highly pathogenic avian influenza A (H5N8) clade 2.3.4.4b viruses in grebes in Inner Mongolia and Ningxia, China, in 2021 | Journal Of Integrative Agriculture | 10.1016/j.jia.2023.09.026 | 2024-01 |
| Yang, Jing | Novel Avian Influenza Virus (H5N1) Clade 2.3.4.4b Reassortants in Migratory Birds, China | Emerging Infectious Diseases | 10.3201/eid2906.221723 | 2023-06 |
| Yang, Qiqi | Synchrony of Bird Migration with Global Dispersal of Avian Influenza Reveals Exposed Bird Orders | Nature Communications | 10.1038/s41467-024-45462-1 | 06/02/2024 |
| Ye, Hejia | Divergent Reassortment and Transmission Dynamics of Highly Pathogenic Avian Influenza A(H5N8) Virus in Birds of China During 2021 | Frontiers In Microbiology | 10.3389/fmicb.2022.913551 | 30/06/2022 |
| Yin, Shenglai | Strong and consistent effects of waterbird composition on HPAI H5 occurrences across Europe | Ecological Applications | 10.1002/eap.3010 | 2024-09 |
| Youk, Sungsu | H5N1 highly pathogenic avian influenza clade 2.3.4.4b in wild and domestic birds: Introductions into the United States and reassortments, December 2021-April 2022 | Virology | 10.1016/j.virol.2023.109860 | 2023-10 |
| Zeng, Jinfeng | Spatiotemporal genotype replacement of H5N8 avian influenza viruses contributed to H5N1 emergence in 2021/2022 panzootic | Journal Of Virology | 10.1128/jvi.01401-23 | 19/03/2024 |
| Zhang, Guogang | Bidirectional Movement of Emerging H5N8 Avian Influenza Viruses Between Europe and Asia via Migratory Birds Since Early 2020 | Molecular Biology And Evolution | 10.1093/molbev/msad019 | 03/02/2023 |
| Zhang, Jiahao | Genomic evolution, transmission dynamics, and pathogenicity of avian influenza A (H5N8) viruses emerging in China, 2020 | Virus Evolution | 10.1093/ve/veab046 | 20/01/2021 |
| Zhang, Jiahao | Survivability of H5N8 mixed wild bird droppings in different conditions | Lancet Microbe | 10.1016/S2666-5247(22)00031-3 | 2022-05 |
| Zhang, Xiaoqing | Highly Pathogenic Avian Influenza A Virus in Wild Migratory Birds, Qinghai Lake, China, 2022 | Emerging Infectious Diseases | 10.3201/eid3010.240460 | 2024-10 |